# Universality, Scaling and Topology with a Modified Lattice Action

J. Fingberg,[a] [*]

[a]SCRI, The Florida State University, Tallahassee, FL 32306-4052, U.S.A.

We investigate universality, scaling, the $\beta$–function and the topological charge in the positive plaquette model for SU(2) lattice gauge theory. The effect of a complete suppression of a lattice artifact, namely the negative plaquettes, on physical quantities, such as the critical temperature, the string tension, the topological charge, glueball masses, and their ratios was examined.

A complete understanding of QCD requires not only the measurement of quantities like the mass spectrum, the critical temperature or the equation of state, but also knowledge about the dynamics of the QCD vacuum. Numerical simulations allow one to study both aspects from first principles. However, it is a well known fact that lattice theories can suffer from artifacts. One example is the famous dip in the step $\beta$-function. It can be linked to a crossover in the behaviour of the average plaquette found for SU(2) pure gauge theory at $\beta \approx 2.2$. The crossover can be seen as a remnant of a line of first order phase transitions of a generalized mixed fundamental-adjoint model in the complex ($\beta$, $\beta_{\text{adjoint}}$)-plane [1]. The same pattern is found for both gauge groups SU(2) and SU(3). A second example is the divergence of the topological suszeptibility $\chi_t$ in the continuum limit. The known methods used to measure $\chi_t$ on the lattice, while equivalent in the classical continuum limit, give different results. On the quantum level rough configurations containing small scale artifacts called dislocations exist in the neighbourhood of smooth ones. These rough configurations can cause the topological suszeptibility to diverge, at least when measured by the geometrical definition [2–4]. Both phenomena are directly connected to a lattice artifact, the negative plaquettes. The crossover in the behaviour of the average plaquette dissapears when negative plaquettes are suppressed [5]. The simplest exceptional configurations that hinder the measurement of the topological charge are those that contain negative plaquettes. One type of such a configuration, the fluxon, has been found to become stable for $C = \beta/\beta_{\text{adjoint}} > 1/2$ [6], which is close to the endpoint of the first order transition line at $C \simeq 0.6$. Indications for a general connection between dislocations on the order of one lattice spacing and negative plaquettes were found by numerically minimizing the action in a given sector of the topological charge [3].

In the continuum region only gauge fields with a small action density contribute to the partition function. The PPM [7–9] is defined by turning this observation into a rigorous constraint: Tr $U_p(x) > 0$. The resulting model was studied with the Wilson plaquette action. First we determined critical couplings from the crossings of the Binder cumulant of the Polyakov loop. We established that the unsubtracted susceptibility (Fig. 1) shows the expected finite-size scaling behaviour [10]. A clear indication for a second order phase transition with critical exponents consistent with those of the standard Wilson formulation was found. At zero temperature we measured the heavy quark potential and determined the $\beta$-function through a Monte Carlo Renormalization Group (MCRG) study. We found a potential that is linearly rising at large distances (Fig. 2) giving a clear signal for confinement. Thus there is hope that the PPM and the Wilson theory describe the same physical situation in the continuum limit. This was confirmed by a test of scaling for ratios of the string tension and glueball masses. We found that physical ra-tios agree and show similar scaling patterns in the PPM and the conventional formulation. In contrast to this similarity we found a strong difference in the behaviour of the step $\beta$-function (Fig. 3). The PPM modification completely removes the dip produced by the SWA. This shows that both features, the peak of the lattice specific heat and the dip of the step $\beta$-function have a common origin, the negative plaquettes, and thus have to be interpreted as lattice artifacts. A comparison of the critical temperature in physical units for the PPM and the SWA (Fig. 4) shows that we are rather far away from the asymptotic scaling region in both cases. However the PPM approaches asymptotic scaling monotonously for the values of the coupling constant under consideration. The rate of convergence is comparably slow in both models. The convergence is vastly improved for both cases if an effective coupling scheme, $\beta_E$ [12], is used. In addition we computed the topological charge with the naive and the geometrical Philipps-Stone algorithm [4] both on hot and cooled [13] configurations. At equal lattice spacing we found that, while the PPM does somewhat better than the SWA, it still contains dislocations. A large drop of the susceptibility between hot and cooled configurations shown in Fig. 5 indicates that it is dominated by short distance dislocations, even in the PPM.

[*]This work was done in collaboration with U.M. Heller and V.K. Mitrjushkin and was supported in part by the DOE under grants # DE-FG05-85ER250000 and # DE-FG05-92ER40742.

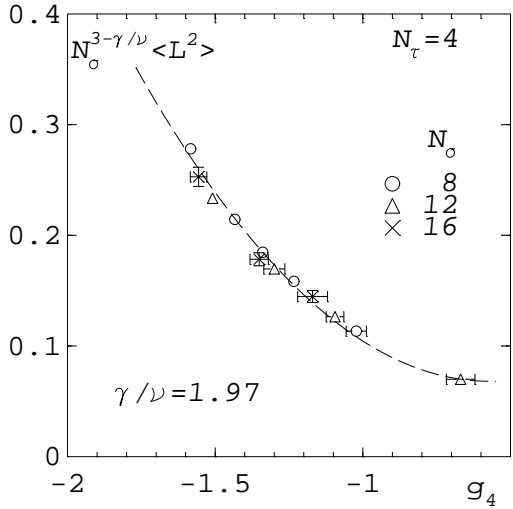

Figure 1. Finite–size scaling of the susceptibility.

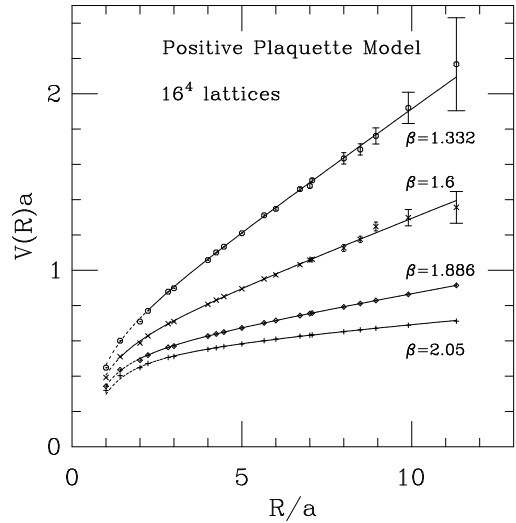

Figure 2. The heavy quark potential.

Our calculations for SU(2) pure gauge theory on symmetric (up to $16^4$) and asymmetric lattices with temporal extent $N_\tau = 2, 4$ and 8 show that the PPM and SWA belong to the same universality class. Physical ratios $m(2^{++}) : m(0^{++}) : \sqrt{\sigma} : \chi_t^{1/4}$ agree and show similar scaling patterns. However, the approach to asymptotic scaling is changed drastically by the PPM modification: the dip in step $\beta$-function is completely removed. While convergence to asymtotic scaling seems to be equally slow we found that effective coupling schemes work equally well for PPM and SWA. In the topological sector, even with fluxons removed, dislocations on larger scales remain. Their identification remains a challenging task on the way to clarifying the fate of the continuum limit of the topological susceptibility.

## REFERENCES

1. G. Bhanot and M. Creutz, Phys. Rev. **D24** (1981) 3212.
2. D.J.R. Pugh and M. Teper, Phys. Lett. **218B** (1989) 326; Phys. Lett. **224B** (1989) 159.
3. M. Göckeler, A.S. Kronfeld, M.L. Laursen, G. Schierholz and U.-J. Wiese, Phys. Lett.

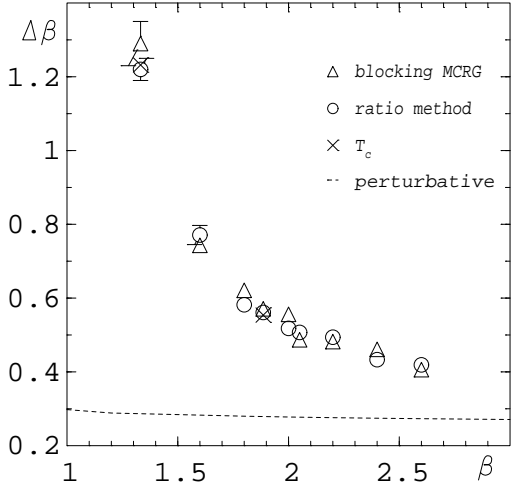

Figure 3. Step $\beta$-function from $T_c$ and MCRG.

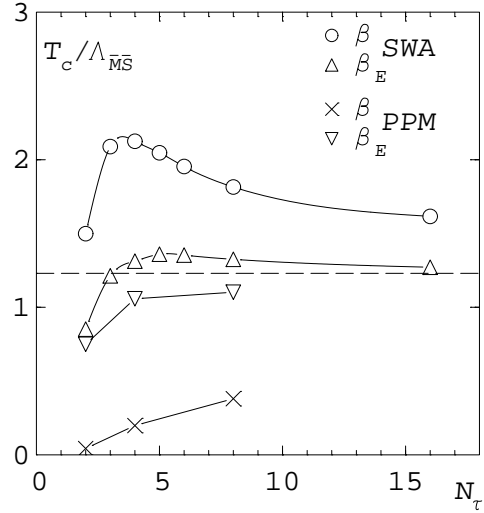

Figure 4. Asymptotic scaling of the critical temperature with bare ($\beta$) and effective coupling ($\beta_E$). The dashed lines indicate the continuum values as determined in Ref. [11].

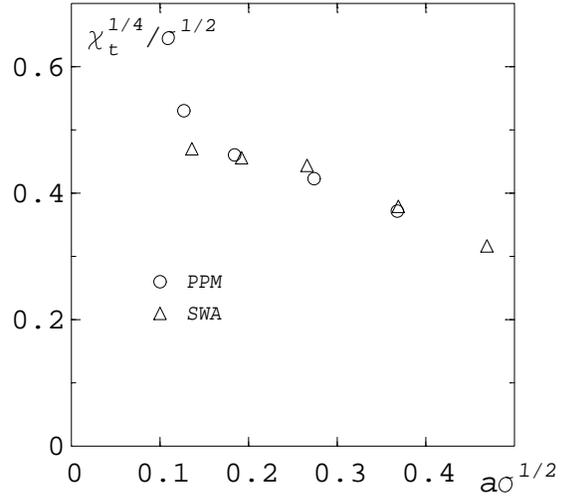

Figure 5. Naive charge rounded away from zero after 15 cooling sweeps on $16^4$ lattices. Results for the SWA are from Ref. [14].

**233B** (1989) 192.
4. A. Philips and D. Stone, Com. Math. Phys. **103** (1986) 599.
5. V.G. Bornyakov, M. Creutz and V. Mitrjushkin, Phys. Rev. **D44** (1991) 3918.
6. R. Dashen, U.M. Heller and H. Neuberger, Nucl. Phys. **B215** (1983) 360.
7. G. Mack and E. Pietarinen, Nucl. Phys. **B205** [**FS5**] (1982) 141.
8. A. Ambjorn and G. Thorleifsson, NBI preprint NBI–HE–94–30, March 1994.
9. U.M. Heller, J. Fingberg and V.K. Mitrjushkin, FSU-SCRI-94-71, to appear in Nucl. Phys. B.
10. J. Engels, J. Fingberg and V.K. Mitrjushkin, Phys. Lett. **B298** (1993) 103.
11. J. Fingberg, U.M. Heller and F. Karsch, Nucl. Phys. **B392** (1993) 493.
12. G. Parisi, in *Proceedings of the $XX^{th}$ Conference on High Energy Physics*, Madison 1980.
13. M. Teper, Phys. Lett. **162B** (1985) 357; **171B** (1986) 81, 86; E.M. Ilgenfritz, M.L. Laursen, M. Müller-Preussker, G. Schierholz and H. Schiller, Nucl. Phys. **B268** (1986) 693.
14. C. Michael and M. Teper, Phys. Lett. **199B** (1987) 95.